\documentclass[reqno]{amsart}
\usepackage{amsmath, amsfonts}
\usepackage{amsthm}
\usepackage{amssymb}
\usepackage{dsfont}
\usepackage{amscd}
\usepackage{color}
\usepackage{graphicx}

\newtheorem{theorem}{Theorem}[section]
\newtheorem{corollary}[theorem]{Corollary}

\newtheorem{prop}[theorem]{Proposition}
\theoremstyle{definition}

\theoremstyle{remark}
\newtheorem{remark}[theorem]{Remark}

\numberwithin{equation}{section}

\begin{document}

\title[Volumes for ${\rm SL}_N(\mathbb R)$]{Volumes for ${\rm SL}_N(\mathbb R)$, the Selberg integral and random lattices}

\author{Peter J. Forrester} \address[Peter J. Forrester]{Department of Mathematics and Statistics, 
ARC Centre of Excellence for Mathematical \& Statistical Frontiers,
The University of Melbourne, Victoria 3010, Australia; 
}\email{pjforr@unimelb.edu.au}

\date{}

\begin{abstract} There is a natural left and right invariant Haar measure associated with the
matrix groups GL${}_N(\mathbb R)$ and SL${}_N(\mathbb R)$ due to Siegel.
For the associated volume to be finite it is necessary to truncate the groups by
imposing a bound on the norm, or in the case of SL${}_N(\mathbb R)$, by restricting
to a fundamental domain. We compute the asymptotic volumes associated with the Haar measure
for GL${}_N(\mathbb R)$ and SL${}_N(\mathbb R)$ matrices in the case of that
the operator norm  lies between $R_1$ and $1/R_2$ in the former, and this norm,
or alternatively the 2-norm, is bounded
by $R$ in the latter. By a result of Duke, Rundnick and Sarnak, such asymptotic
formulas in the case of SL${}_N(\mathbb R)$ imply an asymptotic counting formula
for matrices in SL${}_N(\mathbb Z)$. We discuss too the sampling of SL${}_N(\mathbb R)$ matrices from the truncated sets. By then using lattice reduction to a fundamental domain, we obtain histograms approximating the probability density functions of the lengths and pairwise angles of
shortest length bases vectors in the case $N=2$ and 3, or equivalently of shortest
linearly independent vectors in the corresponding random lattice. In the case $N=2$
these distributions are evaluated explicitly.
\end{abstract}

\maketitle
\section{Introduction}

Fundamental to random matrix theory is the notion of an invariant measure, also
referred to as Haar measure.
For the classical matrix groups SO$(N)$ and U$(N)$ the invariant measure was determined
by Hurwitz \cite{Hu97} in a pioneering paper written in the late 1890's. The recent work
\cite{DF15} documents the importance of this paper as seen from subsequent developments
in random matrix theory.

One place where Hurwitz's idea of an invariant measure on matrix spaces
is pivotal, but which appears to be little known in the
random matrix theory community, is in Siegel's work \cite{Si45} on the geometry of numbers.
In \cite{Si45} Siegel took up the problem of defining an invariant measure on the space of
random unimodular lattices, being guided by both \cite{Hu97} and, according to
\cite{MR58}, the work of Minkowski \cite{Mi05} on the theory of quadratic forms.
The first step in  \cite{Si45} is to define an invariant measure on the matrix
group ${\rm SL}_N(\mathbb R)$ of all $N \times N$ real matrices with unit determinant.
Unlike SO$(N)$ and U$(N)$, this set is not compact, and in particular does not 
have a finite volume. 

In developing the work of Siegel, Macbeath and Rogers \cite{MR55} introduced a
truncation of ${\rm SL}_N(\mathbb R)$, defined by requiring that the operator
norm $|| M||_{\rm Op} := \mu_1$, where $\mu_1$ is the largest singular value of
$M$, be bounded by some value $L$. 
Later Duke, Rudnick and Sarnak \cite{DRS93} considered a similar truncation,
now requiring that the 2-norm $|| M||_2 := ( \sum_{j=1}^N \mu_j^2 )^{1/2}$, where
$\mu_j$ is the $j$-th largest singular value, be bounded.
In \S \ref{S2.1} and \ref{S2.2}
we show that the problem of computing the volume of these sets, discussed in
\cite{JM59} and  \cite{DRS93}  using methods which have not been followed up
in subsequent literature can, alternatively, be approached using integration methods
for matrix integrals in common use in random matrix theory and involving the Selberg
integral \cite{Se44,FW07p}.

Next, in \S \ref{S2.3}, we consider the problem of computing the asymptotic volume of these and similar truncated
sets in the $R \to \infty$ limit. Actually, there are already a number of such computations
in the literature \cite{JM59, DRS93, Ja67}. As pointed out by
Duke, Rudnick and Sarnak \cite{DRS93} these have an arithmetic/ combinatorial significance. 
Thus consider the subgroup ${\rm SL}_N(\mathbb Z)$ of ${\rm SL}_N(\mathbb R)$, so that
the entries of the matrices are now integers. Then we have from
 \cite{DRS93} (see also \cite{GM03}) that
 \begin{equation}\label{Sa}
 \# \{ \gamma: \gamma \in {\rm SL}_N(\mathbb Z), || \gamma || \le R \}
 \mathop{\sim}\limits_{R \to \infty} {1 \over {\rm vol} \, \Gamma}
 \int_{||G|| \le R} (d G),
 \end{equation}
 where $(dG)$ is the Haar measure on ${\rm SL}_N(\mathbb R)$,
 and $ {\rm vol} \, \Gamma$ the volume of a fundamental domain, which
 has the known explicit evaluation in terms of the Riemann zeta
 function (see e.g.~\cite{MR58})
 \begin{equation}\label{Sa1a} 
 {\rm vol} \, \Gamma = \zeta(2) \zeta(3) \cdots \zeta(N). 
 \end{equation} 
  This holds independent
 of the particular norm, provided it is orthogonally invariant. Knowledge of the asymptotic
 form of the RHS of (\ref{Sa}) in the case of $|| \cdot || = || \cdot ||_{\rm Op}$ then gives
 an asymptotic counting formula distinct from that already noted in \cite{DRS93} for
 $|| \cdot || = || \cdot ||_2$.

Other interesting problems show themselves.
One is that of
sampling matrices with invariant measure from the truncated sets, and sampling too
the intersection of these sets with the fundamental domain \cite{Ri16}. From the
latter one can obtain estimates (and analytic formulas for $N=2$) of the distribution
of the corresponding bases vectors of the random lattice. We carry out this study in
Section 4, after computing the averaged characteristic polynomial is Section 3, the
zeros of which can be used as initial conditions in a Metropolis Monte Carlo sampling.

\section{Invariant measure and volumes}
\subsection{${\rm GL}_N(\mathbb R)$}\label{S2.1}

The matrix group ${\rm GL}_N(\mathbb R)$ is the set of all real $N \times N$ invertible 
matrices. Let $(dG)$ denote the product of differentials of the independent entries,
so that for $G = [g_{ij}]_{i,j=1,\dots,N}$, $(dG) = \prod_{i,j=1}^N d g_{i,j}$.
For $A \in {\rm GL}_N(\mathbb R)$ and fixed, one has (see e.g.~\cite{Ma97})
\begin{equation}\label{AG}
(d(AG)) = |\det A|^N (d G), \quad (d(GA)) = |\det A|^N (d G).
\end{equation}
As a consequence
\begin{equation}\label{AG1}
{(d G) \over |\det G|^N}
\end{equation}
is unchanged by both left and right multiplication of $G$ by independent elements in
${\rm GL}_N(\mathbb R)$, and is thus a left and right invariant Haar measure for the group. As mentioned in
the Introduction, such invariant measures were introduced by Hurwitz \cite{Hu97} for
the classical matrix groups SO$(N)$ and U$(N)$. Here, with
$R \in {\rm SO}(N)$ and $U \in {\rm U}(N)$, the analogue of 
(\ref{AG1}) is
$$
(R^T dR) \quad {\rm and} \quad (U^\dagger d U).
$$
Hurwitz  \cite{Hu97} used parameterisations of 
SO$(N)$ and U$(N)$ in terms of Euler angles to obtain explicit formulas for
the invariant measure and from this computed the associated volumes
of these classical group. In distinction to these examples, which are compact sets,
the invariant measure for ${\rm GL}_N(\mathbb R)$ does not have finite volume,
unless the integration is carried out over restricted domains.

Perhaps the most natural restricted domain is specified by 
\begin{equation}\label{AG1a}
D_{R_1, R_2}^{||\cdot||}( {\rm GL}_N(\mathbb R)) := \{ M \in {\rm GL}_N(\mathbb R): R_1 \ge ||M^{-1}|| \; {\rm and} \:
||M|| \le R_2 \},
\end{equation}
where $R_2 R_1 \ge 1$. As remarked in \cite{FR15}, in the context of
selecting elements uniformly at random form  SL${}_N(\mathbb Z)$, this is the case $R_2=1/R_1$ is analogous to bounding the condition number $||M|| \, ||M^{-1}||$.
We would like to compute ${\rm vol} \, D_{R_1, R_2} $, which is defined as
the invariant measure (\ref{AG1}) integrated over  $D_{R_1, R_2}$.
This is tractable for the norm $|| \cdot || = || \cdot ||_{\rm Op}$, when we have
\begin{equation}\label{AG2}
D_{R_1, R_2}^{||\cdot||_{\rm Op}} ({\rm GL}_N(\mathbb R) )= \{ M \in {\rm GL}_N(\mathbb R): 1/R_1 \le \sigma_N \; {\rm and} \:
\sigma_1  \le R_2 \}.
\end{equation}

To compute the volume, as done in  \cite{JM59} in relation to computing a similar volume
in the case of SL${}_N(\mathbb R)$ (see the next subsection), we make use of the singular value decomposition
\begin{equation}\label{MR}
M = O_1 {\rm diag} \, (\sigma_1,\dots, \sigma_N) O_2^T,
\end{equation}
where $O_1, O_2 \in O(N)$ and $\{\sigma_i \}$ are the singular values, ordered
$\sigma_1 \ge \sigma_2 \ge \cdots \sigma_N > 0$. The fact that
$M^T M = O_2 {\rm diag} \, (\sigma_1^2,\dots, \sigma_N^2) O_2^T$ implies that
$\{ \sigma_i^2 \}$ are uniquely determined as the eigenvalues of $M^T M$, while
$O_2$ is the matrix of eigenvectors. For the latter to be uniquely determined we require
that the entries of the first row be positive. Substituting in (\ref{MR}) we see that $R_1$
is uniquely determined and that its image is all of O$(N)$.

The explicit computation of the Jacobian for the change of variables from the elements
of $M$ to variables representing the independent elements on the RHS of (\ref{MR})
was carried out in \cite{JM59}, and with $(dM) := \prod_{i,j=1}^N dM_{i,j}$ one has
\begin{equation}\label{MR1}
(dM) = 2^{-N} (O_1^T dO_1) (O_2^T dO_2) \prod_{1 \le j < k \le N}
(\sigma_j^2 - \sigma_k^2) \, d \sigma_1 \cdots d \sigma_N.
\end{equation}
Here $ (O_1^T dO_1)$ and $ (O_2^T dO_2)$ are the invariant measures on O$(N)$
as identified by  Hurwitz \cite{Hu97}. The factor $2^{-N}$ comes about
due to the restriction on the sign of the first row in $O_2$. An essential point is that
the dependence on $O_1$ and $O_2$ factorises from the dependence on the
eigenvalues. Thus we have
\begin{multline}\label{MR2}
{\rm vol} \, D_{R_1, R_2}^{||\cdot||_{\rm Op}}({\rm GL}_N(\mathbb R) ) =
\\
2^{-N} \Big ( {\rm vol} \, O(N) \Big )^2
\int_{R_1 > \sigma_1 > \cdots > \sigma_N > 1/R_2}
\prod_{l=1}^N \sigma_l^{-N}
 \prod_{1 \le j < k \le N}
(\sigma_j^2 - \sigma_k^2) \, d \sigma_1 \cdots d \sigma_N.
\end{multline}
The value of $ {\rm vol} \, {\rm O}(N)$ was calculated by Hurwitz \cite{Hu97}
(see e.g.~\cite[Th.~2.1.15]{Mu82} and Remark \ref{R2.3} below),
\begin{equation}\label{VO}
{\rm vol} \,({\rm O}(N)) = 2^{N} \prod_{k=1}^N {\pi^{k/2} \over \Gamma(k/2)}.
\end{equation}
 In the limit $R_1R_2 \to \infty$ it also possible
to specify the leading asymptotic form of the integral in (\ref{MR2}).

\begin{prop}\label{p2.1}
Define the PDF on $[0,1]^N$
\begin{equation}\label{MR3}
{1 \over S_N}  \prod_{1 \le j < k \le N} | x_j - x_k|,
\end{equation}
where $S_N$ is the normalisation
(the latter is the case $\lambda_1=\lambda_2=0$, $\lambda=1/2$ of the Selberg
integral, using the notation of \cite[Ch.~4]{Fo10}).
Denote the multidimensional integral in (\ref{MR2}) by $I_N(R_1,R_2)$. This
can be written as an average over the PDF (\ref{MR3}),
\begin{equation}\label{MR4}
I_N(R_1,R_2) = {2^{-N} S_N \over N!} \Big \langle
\prod_{l=1}^N  \Big (1+ 1/(R_1 R_2)^2 - x_l \Big )^{-(N+1)/2} \Big \rangle.
\end{equation}
Introduce the notation $A(x) \asymp B(x)$ to mean that there exists two positive
numbers $C_1$ and $C_2$ independent of $x$ such that
$$
C_1 \le {A(x) \over B(x)} \le C_2.
$$
In the limit $R_1R_2 \to \infty$ we have, for $N$ odd
\begin{equation}\label{MR4o}
I_N(R_1,R_2) \asymp (R_1 R_2)^{(N^2-1)/4} \log (R_1 R_2),
\end{equation}
while for $N$ even
\begin{equation}\label{MR4e}
I_N(R_1,R_2) \asymp (R_1 R_2)^{N^2/4}.
\end{equation}
\end{prop}

\noindent
Proof. \quad The change of variables $\sigma_l^2 = x_l$, $x_l \mapsto x_l + 1/R_2^2$,
$x_l \mapsto R_1^2 x_l$ shows the validity of (\ref{MR4}). The $1/(R_1 R_2)^2
\to 0$
asymptotics of a class of averages including (\ref{MR4}) have been studied in
\cite{FK04, FR11}, and from the results therein we read off (\ref{MR4o}) and
(\ref{MR4e}). \hfill $\square$

\medskip
\begin{remark}
The product of differences in (\ref{MR2}) can be written as a Vandermonde determinant,
which in turn is equivalent to the expression ${\rm Asym} \, \prod_{l=1}^N \sigma_l^{2(N-l)}$.
With $N$ even, if we consider only the diagonal term $\prod_{l=1}^N \sigma_l^{2(N-l)}$,
and integrate to the upper terminal $R_1$ for $\sigma_1,\dots, \sigma_{N/2}$,
and to the lower terminal $R_2$ for $\sigma_{N/2+1}, \dots , \sigma_N$ we reclaim
(\ref{MR4e}). With $N$ odd, (\ref{MR4o}) can be reclaimed by now integrating to
the upper terminal $R_1$ for $\sigma_1,\dots, \sigma_{(N-1)/2}$, to the lower
terminal $R_2$ for $\sigma_{[N+1)/2},\dots, \sigma_{N}$ and between both terminals
for $\sigma_{(N+1)/2}$. Also, direct calculation can be used to evaluate the integral
explicitly for small $N$, and from this we read off that
\begin{equation}\label{RO}
I_2(R_1,R_2) \sim (R_1 R_2), \qquad I_3(R_1,R_2) \sim
{(R_1 R_2)^2 \log R_1 R_2 \over 4},
\end{equation}
which are consistent with (\ref{MR4o}) and (\ref{MR4e}) and furthermore give the
proportionality constants. General formulas for the latter are also given in
\cite{FR11}. For $N=2$ the first result in (\ref{RO}) is reclaimed. For $N=3$ and
beyond ill defined quantities are encountered. In particular, for 
$N=3$ one needs to interpret the quantity $\sin \pi \lambda_1/ 
\sin \pi (\lambda_1 + \alpha)$ in the limit that $\lambda_1 \to 0$ and
$\alpha \to -1$.
\end{remark}

\begin{remark}\label{R2.3}
Hurwitz's evaluation \cite{Hu97} of ${\rm vol} \,({\rm O}(N)) $ actually differs from
(\ref{VO}) by an additional factor of $2^{N(N-1)/4}$. This is due to the 
particular embedding of the space of orthogonal matrices in Euclidean space as chosen
by Hurwitz; see e.g.~\cite[Eq.~(3.10) and surrounding text]{DF15}. To check that
(\ref{VO}) is consistent with (\ref{MR1}) we can multiply both sides by
$ \pi^{-N^2/2} e^{-{\rm Tr} \, M^T M}$ and integrate over $M$. On the LHS we get unity.
On the RHS, after a simple change of variables we obtain
$$
2^{-2N} { ({\rm vol} \, {\rm O}(N))^2 \over N!}
\int_0^\infty \cdots \int_0^\infty \prod_{l=1}^N \sigma_l^{-1/2} e^{- \sigma_l}
\prod_{1 \le j < k \le N} | \sigma_j - \sigma_k | \, d\sigma_1 \cdots
d \sigma_N.
$$
This multidimensional integral is a particular example of a limiting case of the
Selberg integral, and has a well known gamma function evaluation
given explicitly by $\pi^{-N/2} N! \prod_{j=1}^N(\Gamma(j/2))^2$;
see \cite[Prop.~4.7.3]{Fo10}. Making use of (\ref{VO}) shows that the RHS also reduces
to unity.
\end{remark}

\subsection{${\rm SL}_N(\mathbb R)$}\label{S2.2}
Matrices $M \in {\rm GL}_N(\mathbb R)$, with the further requirement that
the determinant is equal to 1, form the 
 group SL${}_N(\mathbb R)$. In \cite{Si45} Siegel considered the associated cone
$\{\lambda M: \: 0 \le \lambda \le 1, \, M \in {\rm SL} _N(\mathbb R) \}$. According to
(\ref{AG1}) the invariant measure
for this cone is simply the Lebesgue measure in $\mathbb R^{N^2}$,
$(d M)$. An equivalent procedure, to be adopted herein, is to impose the
delta function constraint $\delta(1 - \det M)$ in the integrand of the invariant
measure for ${\rm GL}_N^+(\mathbb R)$ (the superscript ``$+$'' here refers to
restricting the determinant to positive values.) In terms of the singular values the
delta function reads $\delta \Big (1 - \prod_{l=1}^N \sigma_l \Big )$.

We take up the problem of computing the volume for the analogue of the domain 
(\ref{AG1a}) in the case of the invariant measure for SL${}_N(\mathbb R)$. 
According to the above remarks, this is given by inserting the delta function
constraint in the integral in (\ref{MR2}), and also dividing by one half due to the
restriction to positive determinant. In distinction to (\ref{AG1a}), this volume
remains finite if we first take $R_2 \to \infty$. Doing this allows us to reduce
the multidimensional integral down to a one-dimensional integral, as first
shown by Jack and Macbeath \cite{JM59}. We give a simplified derivation.

\begin{prop}\label{p2.4}
Let
\begin{equation}\label{J11}
J_N(R) := 
\int_{R > \sigma_1 > \cdots > \sigma_N > 0}
\delta \Big (1 - \prod_{l=1}^N \sigma_l \Big )   \prod_{1 \le j < k \le N}
(\sigma_j^2 - \sigma_k^2) \,  d \sigma_1 \cdots d \sigma_N.
\end{equation}
Let $c > N - 1$ and
\begin{equation}\label{J11a}
B_N =  {2^{N(N-1)/2}  \over N!}
\prod_{j=0}^{N-1} {\Gamma(1+j/2) \Gamma(3/2 + j/2) \over \Gamma(3/2)} .
\end{equation}
With $[\cdot]$ denoting the integer part, we have
\begin{equation}\label{J12}
J_N(R) =   
{B_N \over 2 \pi i}
\int_{c - i \infty}^{c + i \infty}
\Big ( {1 \over w} \Big )^{[(N+1)/2]} {R^{Nw} \over \prod_{r=1}^{N-1} (w^2 - (N-r)^2)^{[(r+1)/2]}}
\, dw.
\end{equation}

\end{prop}

\noindent
Proof. \quad 
Introduce a parameter $t$ by defining
\begin{equation}\label{JR}
J_N(R;t) := 
\int_{R > \sigma_1 > \cdots > \sigma_N > 0}
\delta \Big (t - \prod_{l=1}^N \sigma_l \Big )   \prod_{1 \le j < k \le N}
(\sigma_j^2 - \sigma_k^2) \,  d \sigma_1 \cdots d \sigma_N.
\end{equation}
After a simple change of variables $\sigma_l^2 = x_l$, taking the Mellin transform of
both sides shows
\begin{align}\label{A1}
\int_0^\infty J_N(R;t) t^{s-1} \, dt 
&  =
{2^{-N} \over N!} \int_0^{R^2} d x_1 \cdots \int_0^{R^2} d x_N \,
\prod_{l=1}^N x_l^{s/2-1} \prod_{1 \le j < k \le N} |x_k - x_j| \nonumber \\
&  =
{2^{-N} R^{Ns} R^{N^2 - N} \over N!}
S_N(s/2-1,0,1/2) \nonumber \\
&  = A_N(R)
R^{Ns} \prod_{j=0}^{N-1} {\Gamma((s+ j)/2) \over \Gamma((s+N+1+j)/2)}.
\end{align}
Here use has been made of the notation for the Selberg integral as defined in
\cite[Ch.~4]{Fo10}, and its gamma function evaluation \cite[Eq.~(4.3)]{Fo10}, as well as
the notation 
\begin{equation}\label{J11a}
A_N(R) =  {2^{-N}R^{N^2- N} \over N!}
\prod_{j=0}^{N-1} {\Gamma(1+j/2) \Gamma(3/2 + j/2) \over \Gamma(3/2)} .
\end{equation}
Now taking the inverse Mellin transform to reclaim $J(R;t)$, and setting
$t=1$ gives
\begin{equation}\label{J12a}
J_N(R) = {A_N(R) \over 2 \pi i}
\int_{c - i \infty}^{c + i \infty}
R^{Ns} \prod_{j=0}^{N-1} {\Gamma((s+ j)/2) \over \Gamma((s+N+1+j)/2)} \, ds,
\end{equation}
valid for $c > 0$. Simplifying the ratio of gamma functions using the appropriate recurrence relation,
and changing variables $s+N-1 = w$ gives (\ref{J12}).

\hfill $\square$

\begin{remark}
Evaluating (\ref{J12}) using the residue theorem gives
\begin{equation}\label{J12b}
J_2(R) = {1 \over 2} (R - R^{-1})^2
\end{equation}
and
\begin{equation}\label{J12b}
J_3(R) = {1 \over 24}(R^6 - R^{-6}) - {1 \over 3} (R^3 - R^{-3}) + {3 \over 2} \log R.
\end{equation}
For general $N$ we can write
\begin{equation}\label{J12c}
J_N(R) = 2 A_N(R) G_{N,N}^{0,N} \bigg (
{\{1 - j/2 \}_{j=0}^{N-1} \atop
\{ - {1 \over 2} (N-1+j) \}_{j=0}^{N-1} } \Big | R^{2N} \bigg ),
\end{equation}
where $G_{p,q}^{m,n}$ denotes the Meijer G-function.
\end{remark}

\begin{remark} The delta function constraint in (\ref{JR}) corresponds to the
distribution of a product of scalar random variables. This structure is very
prevalent in exact computations relating to the eigenvalues and singular values
of products of complex random matrices, as is the appearance of the
Meijer G-function; see e.g.~\cite{AB12,AIK13,KKS15}.
\end{remark}

\medskip

To compute the $R \to \infty$ asymptotics of $J_N(R)$ it is most convenient to use the
form (\ref{J12a}).
Closing the contour in the left half plane and
considering the pole resulting from the term $j=0$ in the product shows that for
$R \to \infty$
\begin{equation}\label{J}
J_N(R) = C_N R^{N(N-1)} + {\rm O}(R^{N(N-2)}),
\end{equation}
where
\begin{equation}\label{J1}
C_N = {2 \over 2^{2N} \Gamma(N/2)}
\prod_{j=0}^{N-1} \Big ( {\Gamma(1+j/2) \over \Gamma(3/2)} \Big )
{\Gamma^2((1+j)/2) \over \Gamma((N+1+j)/2)}.
\end{equation}
The large $R$ form of
\begin{equation}\label{J2}
D_R^{||\cdot||_{\rm Op}}({\rm SL}_N(\mathbb R)) : = \{ M \in {\rm SL}_N(\mathbb R): \: \sigma_1  \le R \}
\end{equation}
is now immediate.

\begin{corollary}\label{C2.5}
For large $R$, and with $C_N$ specified by (\ref{J1}),
\begin{align}\label{J3}
{\rm vol} \, D_R^{||\cdot||_{\rm Op}}({\rm SL}_N(\mathbb R)) & = 2^{-N-1} \Big ( {\rm vol} \, O(N) \Big )^2
C_N R^{N(N-1)} +  {\rm O}(R^{N(N-2)}) \nonumber \\
& = {\pi^{N^2/2} \over \Gamma(N/2)}
\prod_{j=0}^{N-1} {\Gamma(1+j/2) \over \Gamma((N+1+j)/2)} R^{N(N-1)}  +  {\rm O}(R^{N(N-2)}) .
\end{align}
\end{corollary}

\noindent
Proof. \quad The first line follows from the analogue of (\ref{MR2}) with the multidimensional
integral therein replaced by ${1 \over 2}J_N(R)$
(the factor of ${1 \over 2}$ is to account for the restriction to a positive determinant), together with (\ref{J}).  The second follows from (\ref{J1}) and (\ref{VO}). \hfill $\square$

\medskip
This is in agreement with \cite{JM59} where this same functional form was deduced,
but without the leading coefficient being evaluated. We remark that in the case
$N=2$ the coefficients evaluate to
\begin{equation}\label{J3a}
C_N  \Big |_{N=2} = {1 \over 2}, \qquad  2^{-N-1} \Big ( {\rm vol} \, O(N) \Big )^2
C_N  \Big |_{N=2} =  \pi^2,
\end{equation}
while for $N=3$ we have
\begin{equation}\label{J3b}
C_N  \Big |_{N=3} = {1 \over 24}, \qquad  2^{-N-1} \Big ( {\rm vol} \, O(N) \Big )^2
C_N  \Big |_{N=3} = {2 \over 3} \pi^4.
\end{equation}

\begin{remark}\label{R2.6}
The domain implied by (\ref{J11}) has been deduced from
(\ref{AG1a}) by taking $R_2 \to \infty$. If instead we set $R_1 = R_2 = R$, the analogue
of (\ref{J11})  reads
\begin{equation}\label{2.24a}
\int_{R > \sigma_1 > \cdots > \sigma_N > 1/R}
\delta \Big (1 - \prod_{l=1}^N \sigma_l \Big )   \prod_{1 \le j < k \le N}
(\sigma_j^2 - \sigma_k^2) \,  d \sigma_1 \cdots d \sigma_N.
\end{equation}
It has been shown by Jack \cite{Ja67} that the leading $R \to \infty$ asymptotics of
this integral is proportional to $R^{[N^2/2]}$.  Interestingly, this is precisely the
asymptotic behaviour as exhibited by the volume of the corresponding set
for GL${}_N(\mathbb R)$ matrices in Proposition \ref{p2.1}, ignoring the logarithm
in (\ref{MR4o}).

The method used in \cite{Ja67} is not able to give the proportionality constants.
In the case $N=2$ an elementary calculation gives this equal to ${1 \over 2}$, as
in (\ref{J3a}). For $N = 3$, the method of the proof of Proposition \ref{p2.4} gives
the task as equivalent to computing the inverse Mellin transform of
$$
I(R;s) = {1 \over 2^3 3!}
\int_{1/R^2}^{R^2} dx_1 \cdots \int_{1/R^2}^{R^2} dx_3 \,
\prod_{l=1}^3 x_l^{s/2-1} \prod_{1 \le j < k \le 3} |x_k - x_j|.
$$
By ordering the variables, and with the help of computer algebra,
this integral can be evaluated explicitly. With this done, computation of
$$
{1 \over 2 \pi i} \int_{c - i \infty}^{c + i \infty} I(R;s) \, ds, \qquad c>0,
$$
by closing the contour in the left half plane shows that the leading large $R$ contribution
comes from the pole at $s=-2$, and that for $R \to \infty$ the leading asymptotic form
is $R^{4}/4$.
\end{remark}

Similar results are also possible in the circumstance that $||\cdot||_{\rm Op}$ is replaced
by $||\cdot||_2$, so that the set under consideration is
\begin{equation}\label{J4}
D_R^{||\cdot||_{2}}({\rm SL}_N(\mathbb R)) : = \{ M \in {\rm SL}_N(\mathbb R): \: 
\sum_{j=1}^N \sigma_j^2  \le R^2 \}.
\end{equation}
The analogue of (\ref{MR2}) for matrices from SL${}_N(\mathbb R)$ is then
\begin{multline}\label{MR2a}
{\rm vol} \, D_{R}^{||\cdot||_2}({\rm SL}_N(\mathbb R) ) =
{1 \over 2^{N +1}} \Big ( {\rm vol} \, O(N) \Big )^2\\
\times {1 \over N!}
\int_{\sigma_l >0: \, \sum_{j=1}^N \sigma_j^2  \le R^2}
\delta \Big (1 - \prod_{l=1}^N \sigma_l \Big )
 \prod_{1 \le j < k \le N}
|\sigma_j^2 - \sigma_k^2| \, d \sigma_1 \cdots d \sigma_N.
\end{multline}
The multidimensional integral in (\ref{MR2a}) can be expressed as a single
contour integral.

\begin{prop}
Denote the multidimensional integral in (\ref{MR2a}), including the
factor of $1/N!$ by $\hat{I}_N(R)$. 
For $c > 0$ we have
\begin{multline}\label{MR2b}
\hat{I}_N(R) = {R^{N(N-1)} \over 2^N N!}
\prod_{j=1}^N {\Gamma(1+j/2) \over \Gamma(3/2)} \\
\times 
{1 \over 2 \pi i}
\int_{c - i \infty}^{c + i \infty} { \prod_{j=1}^N \Gamma(s/2+ (N-j)/2) \over
\Gamma(sN/2 + N(N-1)/2 + 1)} R^{sN} \, ds.
\end{multline}
\end{prop}

\noindent
Proof. \quad Introducing 
$$
K_N(r,t):= {1 \over N!} \int_0^\infty d \sigma_1 \cdots  \int_0^\infty d \sigma_N \,
\delta \Big ( r^2 - \sum_{p=1}^N \sigma_p^2 \Big )
\delta  \Big (t - \prod_{l=1}^N \sigma_l \Big )
\prod_{1 \le j < k \le N}
|\sigma_j^2 - \sigma_k^2|,
$$
we see that
\begin{equation}\label{J5}
\hat{I}_N(R) = 2 \int_0^R K_N(r,t)\Big |_{t=1} r \, dr.
\end{equation}
We note
\begin{multline*}
\int_0^\infty K_N(r,t) t^{s-1} \, dt =
{1 \over 2^N N!}  \\
\times
\int_{\mathbb R_+^N}
\delta \Big ( r^2 - \sum_{p=1}^N x_p \Big )
\prod_{l=1}^N x_l^{s/2 - 1}
\prod_{1 \le j < k \le N} |x_k - x_j| \,
dx_1 \cdots dx_N.
\end{multline*}
The dependence on $r$ can be scaled out of this latter integral to give
\begin{multline}\label{J6}
\int_0^\infty K_N(r,t) t^{s-1} \, dt \\ =
{ r^{N(s+N-1)-2} \over 2^N N!} 
\int_{\mathbb R_+^N}
\delta \Big ( 1 - \sum_{p=1}^N x_p \Big )
\prod_{l=1}^N x_l^{s/2 - 1}
\prod_{1 \le j < k \le N} |x_k - x_j| \,
dx_1 \cdots dx_N.
\end{multline}

The multidimensional integral in (\ref{J6}) is known \cite{ZS01}, 
\cite[Exercises 4.7 q.3]{Fo10} to be closely related to
the Selberg integral, and has the gamma function evaluation (see also Remark \ref{R27}
below)
\begin{equation}\label{J7}
{1 \over \Gamma(sN/2 + N(N-1)/2 )}
\prod_{j=1}^N {\Gamma(s/2 + (N-j)/2) \Gamma(1 + j/2) \over \Gamma(3/2)}.
\end{equation}
Substituting this in (\ref{J6}), and integrating over $r$ as required in (\ref{J5})
we see that
\begin{multline*}
 \int_0^\infty \Big (2  \int_0^R K_N(r,t) r \, dr \Big ) t^{s-1} \, dt \\ =
{R^{N(s+N-1)} \over 2^N N! \Gamma(sN/2 + N(N-1)/2 +1  )}
\prod_{j=1}^N {\Gamma(s/2 + (N-j)/2) \Gamma(1 + j/2) \over \Gamma(3/2)}.
\end{multline*}
Now taking the inverse Mellin transform and setting $t=1$ as required in
(\ref{J5}) gives (\ref{MR2b}).
\hfill $\square$

\begin{remark}\label{R27}
The following working is an alternative to that in \cite{ZS01}, \cite[Exercises 4.7 q.3]{Fo10}
for the evaluation of (\ref{J6}). Define
$$
D_N(t) := \int_{\mathbb R_+^N}
\delta \Big ( t - \sum_{p=1}^N x_p \Big )
\prod_{l=1}^N x_l^{s/2 - 1}
\prod_{1 \le j < k \le N} |x_k - x_j| \,
dx_1 \cdots dx_N.
$$
Taking the Laplace transform of both sides gives
\begin{align*}
\int_0^\infty e^{-\mu t} D_N(t) \, dt & =
 \int_{\mathbb R_+^N} e^{- \mu \sum_{p=1}^N x_p}
 \prod_{l=1}^N x_l^{s/2 - 1} \prod_{1 \le j < k \le N}
 | x_k - x_j| \\
 & = \mu^{-Ns/2 - N(N-1)/2}
 \prod_{j=0}^{N-1} {\Gamma(1 + (j+1)/2) \Gamma(s/2 + 1 + j/2) \over
 \Gamma(3/2)},
 \end{align*}
 where the second line follows by scaling out the dependence on $\mu$, and
 recognising the resulting multidimensional integral as a particular limiting case
 of the Selberg integral, with a known gamma function evaluation
 \cite[Prop.~4.7.3]{Fo10}. Noting that the inverse Laplace transform of
 $\mu^{-p}$ is $t^{p-1}/\Gamma(p)$ we conclude that
 $$
 D_N(t) = { t^{Ns/2 + N(N-1)/2} \over \Gamma(Ns/2 + N(N-1)/2)}
 \prod_{j=0}^{N-1} {\Gamma(1 + (j+1)/2) \Gamma(s/2 + 1 + j/2) \over
 \Gamma(3/2)}.
 $$
 Setting $t=1$ reclaims  (\ref{J7}).
 \end{remark}
 
 \begin{remark}
 For $N=2$, use of the residue theorem permits the integral in (\ref{MR2b}) to
 be evaluated to give
 $$
 \hat{I}_2(R) = {R^2 \over 2} - 1.
 $$
 For general $N$ the integral in  (\ref{MR2b}) can expressed in terms of a
 Meijer G-function, analogous to (\ref{J12c}).
\end{remark}
 
 Closing the contour in (\ref{J6}) in the left half plane we see that for large $R$ the
 pole at $s=0$ gives the leading order contribution. Evaluating the residue shows that
 in this limit
 \begin{equation}\label{I}
\hat{I}_N(R) = \hat{C}_N R^{N(N-1)} + {\rm O}(R^{N(N-2)})
\end{equation}
where
\begin{equation}\label{I1}
\hat{C}_N = {2 \over 2^{2N} \Gamma(N/2)} {1 \over \Gamma(N(N-1)/2+1)}
\prod_{j=1}^{N}  {\Gamma^2(j/2) \over \Gamma(3/2)}.
\end{equation}
The large $R$ form of the volume (\ref{MR2a}) now follows.

\begin{corollary}\label{C2.9}
For large $R$, and with $\hat{C}_N$ specified by (\ref{I1}),
\begin{align}\label{J3}
{\rm vol} \, D_R^{||\cdot||_{2}}({\rm SL}_N(\mathbb R)) &= 2^{-N-1} \Big ( {\rm vol} \, O(N) \Big )^2
\hat{C}_N R^{N(N-1)} +  {\rm O}(R^{N(N-2)}) \nonumber \\
& = { \pi^{N^2/2} \over \Gamma(N/2) \Gamma(N(N-1)/2 + 1)}  R^{N(N-1)} +  {\rm O}(R^{N(N-2)}).
\end{align}
\end{corollary}

\noindent
Proof. \quad The first line  follows from (\ref{MR2a}) with the definition of $\hat{I}_N(R)$,
and the result (\ref{I}).  The second uses (\ref{I1}) and (\ref{VO}). \hfill $\square$

\medskip
An equivalent result, using different methods, has been given in
\cite[Eq.~(A1.15)]{DRS93}. Also,  we remark that in the case
$N=2$ the coefficients evaluate to
\begin{equation}\label{I3a}
\hat{C}_N  \Big |_{N=2} = {1 \over 2}, \qquad  2^{-N-1} \Big ( {\rm vol} \, O(N) \Big )^2
\hat{C}_N  \Big |_{N=2} =  \pi^2,
\end{equation}
while for $N=3$ we have
\begin{equation}\label{I3b}
\hat{C}_N  \Big |_{N=3} = {1 \over 48}, \qquad  2^{-N-1} \Big ( {\rm vol} \, O(N) \Big )^2
\hat{C}_N  \Big |_{N=3} = {1 \over 3} \pi^4.
\end{equation}
According to the definitions $\hat{I}_N(R) < J_N(R)$ and consequently
$\hat{C}_N \le C_N$.  This latter property is illustrated upon comparing
 (\ref{J3a}) and  (\ref{I3a}), and  (\ref{J3b}) and  (\ref{I3b}).
 
 \subsection{Asymptotic counting formulas for matrices in SL${}_N(\mathbb Z)$}\label{S2.3}
The formula (\ref{Sa}) of Duke, Rudnick and Sarnak \cite{DRS93}, combined with
Corollaries \ref{C2.5} and \ref{C2.9}, gives an asymptotic counting formula for
matrices in SL${}_N(\mathbb Z)$, as made explicit in \cite{DRS93} for
$|| \cdot || = || \cdot ||_2$. Our results above extend the latter formula to
include $|| \cdot || = || \cdot ||_{\rm Op}$.

\begin{prop}
Let $|| \cdot || = || \cdot ||_2$ or $|| \cdot || = || \cdot ||_{\rm Op}$.
For large $R$, and with ${\rm vol}\, \Gamma$ given by (\ref{Sa1a}), we have
\begin{equation}\label{6}
\# \{ \gamma: \gamma \in {\rm SL}_N(\mathbb Z), || \gamma || \le R \}
 \mathop{\sim}\limits_{R \to \infty} {   k_N^{|| \cdot||} \over {\rm vol} \, \Gamma}
 R^{N(N-1)},
\end{equation}
where
 \begin{equation}
 k_N^{|| \cdot||_2} =
 { \pi^{N^2/2} \over \Gamma(N/2) \Gamma(N(N-1)/2 + 1)} 
\end{equation}
and
  \begin{equation}
 k_N^{|| \cdot||_{\rm Op}} =  
 {\pi^{N^2/2} \over \Gamma(N/2)}
\prod_{j=0}^{N-1} {\Gamma(1+j/2) \over \Gamma((N+1+j)/2)}.
\end{equation} 
\end{prop}

In view of our knowledge of the large $R$ form of
$
 \int_{ ||G||_{\rm Op}, ||G^{-1}||_{\rm Op} \le R} (d G)$ as noted in Remark \ref{R2.6}, one might
 wonder if
  \begin{equation}
 \# \{ \gamma: \gamma \in {\rm SL}_N(\mathbb Z), (|| \gamma ||_{\rm Op} , ||\gamma^{-1}||_{\rm Op} \le R) \}
 \mathop{\sim}\limits_{R \to \infty}^{?} {1 \over {\rm vol} \, \Gamma}
 \int_{ ||G||_{\rm Op}, ||G^{-1}||_{\rm Op} \le R} (d G).
 \end{equation}
 If true, the result of Jack \cite{Ja67} would give that the leading large $R$ form is
 proportional to $R^{[N^2/2]}$, which for $N > 2$ is distinct from the $R$ dependence
 in   (\ref{6}).
 
 \section{The averaged characteristic polynomial}
 Let $J_N(R)$ be defined by (\ref{J11}). From the Jacobian formula
(\ref{MR1}), the singular values of matrices from
${\rm SL}_N(\mathbb R)$ chosen with invariant measure, and constrained to
have operator norm less than or equal to $R$, have PDF given by
\begin{equation}\label{H1k}
{1 \over J_N(R)} 
\delta \Big (1 - \prod_{l=1}^N \sigma_l \Big )   \prod_{1 \le j < k \le N}
(\sigma_j^2 - \sigma_k^2)
\chi_{R > \sigma_1 > \cdots > \sigma_N > 0}.
\end{equation} 
Information on a typical sample from this PDF can be obtained from the zeros of
the averaged characteristic polynomial. Integration methods used in \S \ref{S2.2}
allow for a specification of this polynomial in terms of certain inverse Mellin transforms.

\begin{prop}\label{p3.1}
Let $p_N(x)$ denote the average characteristic polynomial for the squared singular 
values of the ensemble (\ref{H1k}), so that
\begin{equation}\label{pN}
p_N(x) := \Big \langle \prod_{l=1}^N(x - \sigma_l^2) \Big \rangle.
\end{equation}
Let ${}_2 F_1$ denote the Gauss hypergeometric function, and suppose $c>0$.
With 
$$
\tilde{J}_N(R) = {1 \over 2 \pi i} \int_{c - i \infty}^{c+i \infty}
R^{Ns}
\prod_{j=0}^{N-1} {\Gamma((s+j)/2) \over \Gamma((s+N+1+j)/2)} \, ds,
$$
we have
\begin{multline}\label{3.3}
p_N(x)  = {(-1)^N \over \tilde{J}_N(R)} \, {1 \over 2 \pi i}  \int_{c - i \infty}^{c+i \infty}
R^{N(s+2)}
\prod_{j=0}^{N-1} {\Gamma((s+j)/2+1) \over \Gamma((s+N+1+j)/2+1)} \\ \times
\, {}_2 F_1(-N,N+s+1;s;x/R^2) \, ds.
\end{multline}
Equivalently, writing (\ref{pN}) as $p_N(x)  =  \sum_{k=0}^N c_k x^k$, we have
\begin{multline}\label{3.4}
c_k =  {(-1)^{N-k}  \over  R^{2k} \tilde{J}_N(R)} \binom{N}{k}\\
\times
{1 \over 2 \pi i}  \int_{c - i \infty}^{c+i \infty}
R^{N(s+2)}
{\Gamma(s) \Gamma(N+s+1+k) \over \Gamma(s+k) \Gamma(N+s+1)}
\prod_{j=0}^{N-1} {\Gamma((s+j)/2+1) \over \Gamma((s+N+1+j)/2+1)} \, ds
\\
= {(-1)^{N-k} \over R^{2(k-N)} }  \binom{N}{k}\
G_{N+4,N+4}^{0,N+4} \bigg ({ \{- j/2  \}_{j=-2}^{N-1},
 \{ - (N-1+k)/2 ,-(N+k)/2 \}
 \atop \{ - (N+1+j)/2 \}_{j=-2}^{N-1}, \{-k/2+1, - (k-1)/2 \} }
\Big | R^{2N} \bigg ) \\
\bigg /
G_{N,N}^{0,N}  \bigg ({ \{- (j/2 - 1) \}_{j=0}^{N-1} \atop \{ - (N-1+j)/2 \}_{j=0}^{N-1} }
\Big | R^{2N} \bigg )
\end{multline}
\end{prop}

\noindent
Proof. \quad  We begin by introducing a parameter $t$ in the delta function as in
(\ref{JR}). Denote the corresponding  averaged characteristic polynomial by 
$p_N(x;t)$. We have
$$
\int_0^\infty p_N(x;t)  t^{s-1} \, dt =
{C_{N,s} \over J_N(R)}  {2^{-N} R^{N(s+2) + N^2 - N} \over N!}
\Big \langle \prod_{l=1}^N(x - x_l/R^2) \Big \rangle,
$$
where the average herein is with respect to the PDF on $[0,1]^N$
$$
{1 \over C_{N,s}} \prod_{l=1}^N x_l^{s/2-1} \prod_{1 \le j < k \le N}
| x_k - x_j|.
$$
According to \cite[Exercises 13.1 q.2]{Fo10} this average is given in terms of the
${}_2 F_1$ function as being equal to
$$
(-1)^N  {C_{N,s+1} \over C_{N,s}} \, {}_2 F_1(-N,N+s+1;s;x/R^2).
$$
Inserting the value of $C_{N,s}$, which is the particular example of the Selberg integral
appearing in (\ref{A1}) and given by the product of gamma functions therein,
the expression (\ref{3.3}) results upon taking the inverse Mellin transform and setting
$t=1$. The explicit form (\ref{3.4}) of the coefficients in the polynomial now follows
by substituting the power series form of the ${}_2F_1$ function in (\ref{3.3}), and
making use of the definition of the Meijer G-function.
\hfill $\square$

\medskip

For a given value of $R$, and values of $N$ up to around 15, the ratio of
Meijer G-functions in (\ref{3.4}) can be evaluated to high accuracy using computer
algebra, and the zeros of $p_N(x)$ computed. For example, with $R=2$ and $N=6$
we find that the zeros occur at
$$
0.04436, \: 0.57774, \: 1.41726, \: 2.33579, \: 3.15342, \: 3.73701.
$$
These are all inside the support $[0,R^2]$ of the squared singular values, and 
furthermore multiply to unity.
It is well known in random matrix theory that the zeros of the characteristic polynomial
are closely related to the spectral density, in the sense that for a broad range
of circumstances it can be proved that
both share the same
density function for large $N$ \cite{Ha15}, although no such theorem is known
in the present setting. Our specific interest in their values will be
as initial conditions for Metropolis Monte Carlo sampling of the PDF
(\ref{H1k}), which we turn to next.

\section{Sampling the invariant measure with applications to random lattices}
\subsection{Sampling from ${\rm SL}_N(\mathbb R)$ 
with bounded norm}
The factorisation of the eigenvector dependence in the Jacobian (\ref{MR1}) for the
singular value decomposition (\ref{MR}) implies that the task of sampling matrices $M$ with
invariant measure and bounded norm from ${\rm SL}_N(\mathbb R)$ reduces to sampling
from the PDF for the singular values. According to (\ref{MR1}) this has the functional form
(\ref{H1k}), further restricted so that $||M|| \le R$.

In the case $N=2$, by integrating out $\sigma_2$ a function of a single variable results.
Explicitly, one obtains
\begin{equation}\label{Dp}
{1 \over C_{2,R}^{||\cdot||}}  {1 \over \sigma_1} \Big (\sigma_1^2 -
{1 \over \sigma_1^2}\Big )
\chi_{||M|| \le R} \chi_{\sigma_1 > 1},
\end{equation}
where $ C_{2,R}^{||\cdot||}$ denotes the normalisation constant.
For $|| \cdot || = || \cdot ||_{\rm Op}$ we have 
$\chi_{||M||\le R} = \chi_{\sigma_1 < R}$, while for $|| \cdot || = || \cdot ||_2$ we have
$\chi_{||M||\le R} = \chi_{\sigma_1 < \tilde{R}}$, where $\tilde{R}^2 =
{1 \over 2} (R^2 + \sqrt{R^4 - 4})$. Thus, up to the precise value of $R$, the same
PDF applies for both norms. For definiteness, let us choose $|| \cdot || = || \cdot ||_{\rm Op}$.
The cumulative distribution is then
\begin{equation}\label{Dp1}
{1 \over C_{2,R}^{||\cdot||_{\rm Op}}} \int_1^r {1 \over \sigma_1}
\Big (\sigma_1^2 - {1 \over \sigma_1^2} \Big ) \, d \sigma_1 =
{(r - 1/r)^2 \over (R-1/R)^2},
\end{equation}
as is consistent with (\ref{J12b}).
Knowledge of this result allows a prescription for the sampling from
the PDF (\ref{Dp}) to be given.

\begin{prop}\label{p4.1}
Let $s$ be a random variable uniformly distributed between 0 and 1. The
random variable
\begin{equation}\label{Dp2}
r = {(R - 1/R) \sqrt{s} + ((R - 1/R)^2 s + 4)^{1/2} \over 2}, \qquad 1 < r < R,
\end{equation}
is distributed according to the PDF  (\ref{Dp}).
\end{prop}

\noindent
{\rm Proof.} \quad This follows by equating (\ref{Dp1}) to $s$ and solving for $r$ as
a function of $s$. \hfill $\square$

\medskip
For ${\rm SL}_N(\mathbb R)$ with $N > 2$ the most straightforward approach to 
sampling the PDF for the distribution of singular values is to adopt a statistical
mechanics viewpoint by writing
$$
\prod_{1 \le j < k \le N} (\sigma_j^2 - \sigma_k^2) =
e^{-E(\{\sigma_l\})}, \qquad
E(\{\sigma_l\}) := - \sum_{1 \le j < k \le N} \log |\sigma_j^2 - \sigma_k^2|,
$$
and to implement the Metropolis Monte Carlo algorithm. However, the situation is not
standard in that all configurations must satisfy the constraint
\begin{equation}\label{SR}
\prod_{l=1}^N \sigma_l = 1, \qquad R > \sigma_l > 0 \: \: (l=1,\dots,N).
\end{equation}
Viewed as a condition on $\sigma_N$, integrating over this variable gives the PDF
for $\{\sigma_l\}_{l=1}^{N-1}$ as 
\begin{equation}\label{SRa}
\sigma_N e^{-E(\{\sigma_l\})} \Big |_{\sigma_N = 1/\prod_{l=1}^{N-1}
\sigma_l} \chi_{ R > \sigma_{1} > \cdots > \sigma_N > 0}
\end{equation}

An initial configuration satisfying (\ref{SR}), which as discussed is expected to well
represent a typical configuration, is given by the zeros of the characteristic
polynomial in Proposition \ref{p3.1}. However, as already commented, for practical purposes their
computation is restricted to values of $N$ up to around 15. For larger $N$
an initial configuration satisfying (\ref{SR}) can be constructed by first forming a vector
of random variables $(x_1,\dots,x_N)$ where $x_j = y_j/\sum_{l=1}^N y_l$ with
each $y_j$ chosen independently from Exp$(1)$. According to a realisation of the
Dirichlet distribution (see e.g.~\cite[Prop.~4.2.4]{Fo10}) this construction implies the $x_j$'s
are uniformly distributed on $[0,1]$ subject to the constraint $\sum_{j=1}^N x_j = 1$.
Next define $X_j = ((x_j - 1/N)/(1 - 1/N)) \log R$ ($j=1,\dots,N)$ so that
$$
\prod_{l=1}^N e^{X_l} = 1, \qquad R > e^{X_l} > 0 \: (l=1,\dots,N).
$$
These facts together imply that by choosing $\sigma_l = e^{X_l}$ $(l=1,\dots,N)$, the constraints (\ref{SR}) are
satisfied. We further order these variables so that $R > \sigma_1 > \cdots > \sigma_N > 0$.

From such an initial condition, or more generally a trial configuration $\{\sigma_l\}$,
an updated configuration $\{\tilde{\sigma}_l\}$ is proposed by picking uniformly
at random a $\sigma_j$ ($j=1,\dots,N-1$), perturbing it by the
rule
$\tilde{\sigma}_j = \sigma_j + \gamma$ and further setting
$\tilde{\sigma}_l = {\sigma}_l$ for $l \ne j,N$
and $\tilde{\sigma}_N = 1/\prod_{l=1}^{N-1} \tilde{\sigma}_l$. Here
$\gamma$ is chosen as a Gaussian random variable with  mean zero and a standard deviation so that the average
rejection rate (see below) is approximately 50$\%$, in accordance with textbook advice relating to the Metropolis algorithm.

\begin{figure}[t]
 \includegraphics{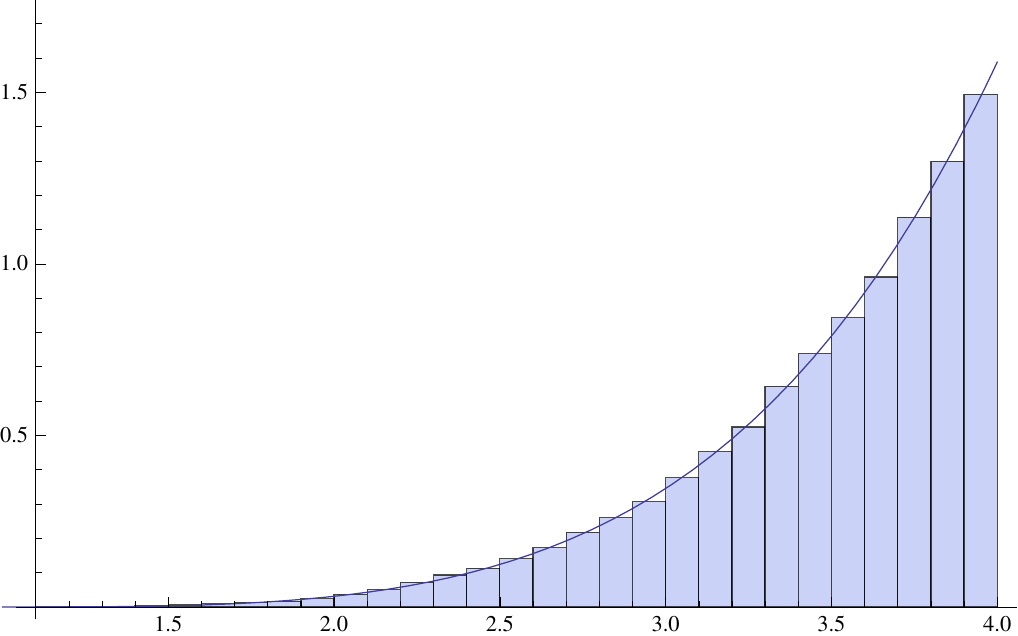} 
 \caption{\label{F1}The distribution of the largest singular value
 as sampled from the PDF (\ref{SRa}) using
 the Metropolis algorithm with $5 \times 10^5$ steps, compared against the
 theoretical value (\ref{yt}). Here $R=4$.}
\end{figure}

The proposed configuration $\{\tilde{\sigma}_l\}$ is immediately rejected if the ordering $R > \tilde{\sigma}_1 > \cdots > \tilde{\sigma}_N > 0$ is violated, and the previous configuration is repeated.
Otherwise one implements the
Metropolis-Hastings rule that the configuration is rejected, and thus the previous
configuration is repeated, with probability $1-p$, where
\begin{equation}\label{p}
p = {\rm min} \, \Big (
{\tilde{\sigma}_N \over \sigma_N} e^{- (E(\{\tilde{\sigma}_l\})  - E(\{\sigma_l\})}, 1 \Big )
\end{equation}
(the factor $\tilde{\sigma}_N /\sigma_N$  results from implementing the delta
function constraint as in (\ref{SRa})).

In the case $N=3$ a test on this methodology is to use it to estimate the distribution
of the largest singular value $\sigma_1$. According to the definition (\ref{J11}) and
(\ref{J12b}), the probability density function for $\sigma_1$, $p_3(s)$ say, is given by
\begin{equation}\label{yt}
p_3(s) = {d \over d s} {J_3(s) \over J_s(R)} =
{ {1 \over 4} (s^5 + s^{-7}) -  (s^2 + s^{-4}) + {3 \over 2 s} \over
{1 \over 24}(R^6 - R^{-6}) - {1 \over 3} (R^3 - R^{-3}) + {3 \over 2} \log R},
\end{equation}
for $R > s > 1$, and $p_3(s) = 0$ otherwise. This test was carried out (using $\gamma = {\rm N}[0,1]$ in the update, and choosing $R=4$), and excellent
agreement  found as exhibited in Figure \ref{F1}.

\subsection{Random lattices}

Matrices in $M \in {\rm SL}_N(\mathbb R)$ relate to lattices. 
To see this, one adopts a viewpoint common in linear algebra
that the columns of $M$ are to be regarded as
vectors in $\mathbb R^N$, denoted $\vec{m}_1, \vec{m}_2,\dots, \vec{m}_N$ say.
Associated with the vectors $\{ \vec{m}_j \}_{j=1,\dots,N}$ is the lattice
$$
\Big  \{ \vec{y}: \vec{y} = \sum_{j=1}^N n_j \vec{m}_j, \quad n_j \in \mathbb Z \: (j=1,\dots,N)
\Big  \}.
$$
Equivalently, $M$ specifies a unit cell of the lattice
\begin{equation}\label{P}
\Big \{ \vec{x}: \vec{x} = \sum_{j=1}^N \alpha_j \vec{m}_j, \quad
0 \le \alpha_j \le 1 \Big  \}.
\end{equation}
Due to the requirement that $\det M = 1$, this has unit volume.

An important point is that
matrices of the form $M \Lambda$ for $\Lambda \in {\rm SL}_N(\mathbb Z)$
(i.e.~the set of $N \times N$ matrices with unit determinant and integer coefficients)
generate the same lattice, and moreover it is easy to verify that for a matrix
$M' \in {\rm SL}_N(\mathbb R)$ to generate the same lattice as $M$, it must be that
there is a $\Lambda \in {\rm SL}_N^\pm(\mathbb Z)$ (we use this notation
for the set of $N \times N$ matrices with integer coefficients and determinant $\pm 1$)
such that $M' = M \Lambda$.
Attention is thus drawn to the quotient space
 ${\rm SL}_N(\mathbb R) / {\rm SL}_N(\mathbb Z) $, which is to be thought of as the space of unimodular lattices.
 
 Crucial to the understanding of $ {\rm SL}_N(\mathbb R)/
{\rm SL}_N(\mathbb Z)  $ is the notion of a fundamental domain
$F \subset {\rm SL}_N(\mathbb R)$. Such a domain ($F$ is not unique) has the
defining properties that ${\rm SL}_N(\mathbb R) = \cup_{\Lambda \in
{\rm SL}_N(\mathbb Z)} F \Lambda$ and also $F \Lambda \cap F$ is empty
for $\Lambda$ not equal to the identity. It follows that up to possible boundary
points $F$ is isomorphic to the quotient space itself.

One way to specify a fundamental domain relates in an essential way to choosing a
distinguished basis for the underlying lattice. Following \cite{NS04}, the qualities one
is seeking is to choose a basis made of reasonably short vectors which are almost
orthogonal. In particular, a basis $\{\mathbf b_1, \dots, \mathbf b_N\}$ is said to be
Minkowski reduced if for all $1 \le i \le N$, $\mathbf b_i$ has minimal norm among
all lattice vectors $\mathbf v$ such that $\{\mathbf b_1,\dots,\mathbf b_{i-1},
\mathbf v \}$ can be extended to a basis. In this definition, the dimensions $N \le 4$ are special:
only then is it that the length of $\mathbf b_i$ must coincide with the so-called
$i$-th minimum, defined as  the radius
of the smallest closed ball centred at the origin and containing $i$ or more
linearly independent lattice vectors.

For $N=2$ it is almost immediate that $\{\mathbf b_1,\mathbf b_2\}$
is Minkowski reduced if 
 \begin{equation}\label{bb}
|| \mathbf b_2|| \ge || \mathbf b_1 ||, \quad 2 | \mathbf b_1 \cdot \mathbf b_2 | \le
 || \mathbf b_1 ||^2, 
 \end{equation}
 as the second inequality is equivalent to requiring that $||
 \mathbf b_2 + n \mathbf b_1 || \ge || \mathbf b_2 ||$ for all $n \in \mathbb Z$.
 For $N = 3$, the definition of a Minkowski  reduced basis in terms of the 
$i$-th minimum  inequalities reads
  \begin{equation}\label{bb1}
|| \mathbf b_3|| \ge || \mathbf b_2|| \ge || \mathbf b_1 ||, \quad
||
 \mathbf b_2 + n_1 \mathbf b_1 || \ge || \mathbf b_2 ||, \quad
 ||
 \mathbf b_3 + n_2 \mathbf b_2 + n_1 \mathbf b_1 || \ge || \mathbf b_3 ||
 \end{equation}
 for all $n_1, n_2 \in \mathbb Z$. 
 
 A natural question is to specify the distributions of the lengths of the Minkowski
 reduced lattice vectors, and/or the first $k$ linearly independent shortest lattice vectors,
  as well as  the angles between them when the lattice is chosen
 at random in the sense that the matrix of basis vectors is an element of
 ${\rm SL}_N(\mathbb R)$ with Haar measure. By using our ability to sample
 the latter (when restricted to have bounded norm) we will show in the cases
 $N=3$ these
 distributions can be approximated by combining the sampling with a lattice reduction
 algorithm \cite{Se01}. In the case $N=2$ analytical calculations are possible, and
 uniform sampling together with the Lagrange--Gauss algorithm for two-dimensional
 lattice reduction can be used to illustrate the results.
 We will take up this task first, before presenting our results for $N=3$. We 
 conclude with a brief discussion of the situation in the $N \to \infty$ limit.
 
 \subsection{The case $N=2$}\label{sS2}
 With $N=2$ the Haar measure for  ${\rm SL}_N(\mathbb R)$ can be parametrised
 in terms of variables simply related to the inequalities (\ref{bb}). One first notes that for
 general $N$, each $M \in {\rm SL}_N(\mathbb R)$ can be decomposed
 $M = QR$, where $Q$ is a real orthogonal matrix with determinant $+1$ and
 $R$ is an upper triangular matrix with diagonal entries all positive. This decomposition is a matrix form of the Gram-Schmidt algorithm reducing the columns of $M$ to an orthonormal basis. From the viewpoint of the space of unimodular lattices, $Q$
 acts as a rotation, and this does not alter the lengths of the reduced lattice vectors
 or the angles between them.
 It is well known in random matrix theory
  \cite{Mu82,Ma97,ER05} that the volume element for the change of variables
  from the elements of $M$ to $Q$ and $R$ is
 \begin{equation}\label{Sa1}
(dM) = \prod_{l=1}^N r_{ll}^{N-l} (dR) (Q^T d Q),
\end{equation}
where $(Q^T dQ)$ is the invariant measure on SO$(N)$ as identified by Hurwitz \cite{Hu97}.

In the case $N=2$ we have
 \begin{equation}\label{R1}
 R = \begin{bmatrix} r_{11} & r_{12} \\ 0 & r_{22} \end{bmatrix}, \qquad r_{22} =
 1/r_{11}.
 \end{equation}
 With the lattice rotated so that $\mathbf b_1$ is chose to lie along the
 positive $x$-axis, we see from (\ref{R1}) that $\mathbf b_1 = (r_{11}, 0)$ and
 $\mathbf b_2 = (r_{12}, r_{22})$, and thus the inequalities (\ref{bb}) read
 $$
 r_{12}^2 + r_{22}^2 \ge r_{11}^2, \qquad 2 | r_{12}| \le r_{11}.
 $$
 From (\ref{Sa1}) and the fact that for $N=2$ we have $\int (Q^T d Q) = 2 \pi$,
 as follows from (\ref{VO}) multiplied by $1/2$ to account for $Q \in {\rm SO}(N)$,
 the volume element of the variables $\{r_{11}, r_{12}, r_{22} \}$ is thus seen to
 be equal to
 $$
 2 \pi \chi_{ r_{12}^2 + r_{22}^2 \ge r_{11}^2}
 \chi_{ 2|r_{12}| \le r_{11}} r_{11} \delta(1 - r_{11} r_{22}) \,
 dr_{11} dr_{12} dr_{22}.
 $$
 After integration over $r_{22}$ this reduces to
 \begin{equation}\label{R2} 
2 \pi  \chi_{r_{11}/2 \ge |r_{12}| \ge A_{r_{11}}(r_{11}^2 - 1/r_{11}^2)^{1/2}}
 dr_{11} dr_{12} ,
  \end{equation}
where $A_r = 1$ for $r \ge 1$, and $A_r = 0$ otherwise. The sought statistical
data can now readily be computed.

\begin{prop}\label{p4.2}
Let {\rm vol}$\, \tilde{\Gamma}$ denote the volume corresponding to
(\ref{R2}). We have
 \begin{equation}\label{R2a} 
 {\rm vol} \,  \tilde{\Gamma} = {\pi^2 \over 3}.
  \end{equation}
  The probability density function
   of the length of the shortest lattice vector $\mathbf b_1$ is given
  by
  \begin{equation}\label{R3}  
  {12 \over \pi}
   \Big ( {s \over 2} - \chi_{s>1} (s^2 - 1/s^2)^{1/2} \Big ), \qquad 0  < s < (4/3)^{1/4}.
  \end{equation}
  The probability density function of the second shortest basis vector
  $\mathbf b_2$ is given by
    \begin{equation}\label{R4}  
  {12 \over \pi s} \Big ( (s^4 -1)^{1/2} \chi_{1<s<(4/3)^{1/4} }+
  (2s^2(s^2-(s^4-1)^{1/2})-1)^{1/2} ) \chi_{(4/3)^{1/4} < s < \infty} \Big ).
  \end{equation}
  The probability density function of  $\cos \theta$, where $\theta$ is the
  angle between $\mathbf b_1$ and $\mathbf b_2$ is
    \begin{equation}\label{R5}  
   - {3 \over 2 \pi} { \log (4 s^2) \over (1 - s^2)^{1/2}}, \qquad 0 < |s| < 1/2.
    \end{equation} 
  
  \end{prop}
  
  \noindent
  Proof. \quad The inequality in (\ref{R2}) tells us that the maximum value of $r_{11}$
  occurs when $r_{11}/2 = (r_{11}^2 - 1/r_{11}^2)^{1/2}$ and thus $r_{11}^4 = 3/4$.
  Using this fact, it follows that
  $$
 {\rm vol} \,  \tilde{\Gamma} = 4 \pi \Big (
 \int_1^{(4/3)^{1/4}} \Big ( {r \over 2} - (r^2 - 1/r^2)^{1/2} \Big ) \, dr + \int_0^1 {r \over 2} \, dr \Big ).
 $$
 Evaluating the integrals gives (\ref{R2a}).
 
 For the distribution of the length of the shortest vector, we know from the text below
 (\ref{R1}) that this length is equal to $r_{11}$. Integrating (\ref{R2}) over
 $r_{12}$, and normalising using    (\ref{R2a}), we obtain (\ref{R3}).
 
 According to the  text below
 (\ref{R1})  the length of the second shortest linearly independent vector is equal to $(r_{12}^2 + 1/r_{11}^2)^{1/2}$.
 Setting this equal to $s$, the inequalities in (\ref{R2}) require that
 $1/s < r_{11} < \sqrt{2}(s^2 - (s^4 - 1)^{1/2})^{1/2}$, while
 $d r_{12} = (t/ r_{12}) dt$. Thus, after changing variables from $r_{12}$ to $s$ in
  (\ref{R1}), our task is compute
  $$
  \int_{1/s}^{ \sqrt{2}(s^2 - (s^4 - 1)^{1/2})^{1/2}}
  {s r \over (r^2 s^2 - 1)^{1/2}} \, dr.
  $$
Doing this and normalising gives (\ref{R4}).

The text below
 (\ref{R1})  tells us that $\cos \theta = r_{12}/(r_{12}^2 + (1/ r_{11})^2)^{1/2}$. Denoting
 this by $s$, the inequalities in (\ref{R2}) require that 
 $(4s^2/(1 - s^2))^{1/4} < r_{11} < 1/(1-s^2)^{1/4}$ and $0 < |s| < 1/2$. Also,
 $dr_{12} = 1/(r_{11}(1-s^2)^{3/2}) \, ds$. Thus, after changing variables from $r_{12}$ to $s$ in
  (\ref{R1}), our remaining task is to compute
  $$
\int_{(4s^2/(1 - s^2))^{1/4}}^{1/(1-s^2)^{1/4}} {1 \over r} \, dr.
$$
Doing this, and after appropriate normalisation, (\ref{R5}) results. \hfill $\square$

\begin{remark}\label{Rm1}
The volume (\ref{R2a}) is equal to twice the value of
${\rm vol} \, \Gamma$ in the case $N=2$ as given by (\ref{Sa}). This can be understood
due to (\ref{Sa}) relating to the fundamental domain of the quotient ${\rm SL}_N(\mathbb R)/ {\rm SL}_N(\mathbb Z)$, whereas in (\ref{R2a})  the quotient is
${\rm SL}_N(\mathbb R)/ {\rm SL}_N^\pm(\mathbb Z)$, where ${\rm SL}_N^\pm(\mathbb Z)$ is the set of all $N \times N$ matrices with integer entries and determinant equal
to $\pm 1$.
\end{remark}

\begin{remark}\label{Rm2}
According to (\ref{R3}) the maximum allowed value of the length of the
shortest vector is $(4/3)^{1/4}$. Suppose that the other basis vector also has this
length. Then, for the resulting unit cell to have area unity, the angle between
the two vectors must be $\pi/3$ or $4 \pi/3$ and so the cosine of the angle must
be $\pm 1/2$, which is the largest value in magnitude permitted by (\ref{R5}).
This corresponds to the triangular, or equivalently hexagonal, lattice.
\end{remark}

\begin{remark}\label{Rm2a}
Consider a punctured disk of radius $0<R<1$ about the origin. According to 
Proposition \ref{p4.2} this disk will contain only the shortest lattice vector and integer
multiples $\pm \mathbf b_1, \pm 2  \mathbf b_1,\dots, \pm  m \mathbf b_1$, where
$m || \mathbf b_1 || < R \le (m+1) || \mathbf b_1 ||$, or equivalently $m = \lfloor R/||\mathbf b_1||
\rfloor$. Thus, with $\Omega (R)$ denoting the expected number of lattice vectors in this
punctured disk, making use of (\ref{R3}) shows 
\begin{equation}\label{w1}
\Omega(R) = {12 \over \pi} \int_0^R \Big \lfloor {R \over s} \Big \rfloor s \, ds =
 {12 R^2 \over \pi} \int_0^1 \Big \lfloor {1 \over s} \Big \rfloor s \, ds.
 \end{equation}
 The latter integral can be written as a sum and evaluated according to
\begin{equation}\label{w2} 
   \int_0^1 \Big \lfloor {1 \over s} \Big \rfloor s \, ds =
   \sum_{p=1}^\infty p \int_{1/(1+p)}^{1/p} r \, dr = 
{1 \over 2} \sum_{p=1}^\infty {1 \over p^2}   =
   {\pi^2 \over 12},
 \end{equation} 
   where the second equality follows by evaluating the integral and simple manipulation of  the resulting summation.
   Hence, for $R < 1$, $\Omega(R) = \pi R^2$, which
   is the area of the corresponding disk. This result, which remains valid for all $R>0$, is a well known consequence of Siegel's mean
   value theorem for lattices; for a readable account see \cite{Pa08}.
   \end{remark}
   
   \begin{remark}\label{R4.6}
  Integration over the invariant measure for SL${}_2(\mathbb R)/{\rm SL}_2(\mathbb Z)$ has been
  carried out in the recent work \cite{MS13} to obtain the explicit functional form  of
 the  distribution of certain scaled diameters for random $2k$-regular circulant graphs with $k=2$. A number of the required integrals had earlier appeared in the works
 \cite{MS08} and \cite{SV05}. Our (\ref{R3}) in fact has an interpretation in the context
 of \cite{SV05}, which relates to the asymptotics of certain random linear congruences
 mod $p$, as $p \to \infty$.  Specifically, it gives the explicit value of $c_1$ in the special
 case $n=2$, $\Omega$ a disk centred at the origin of \cite[Theorem 2]{SV05}, while 
 restricting the radius of the disk to less than 1,
 our Remark \ref{Rm2a} implicitly contains the formula for $c_3,c_5,\dots$ as well (each
 $c_{2j}$ vanishes by symmetry).
 In \cite[Prop.~3]{SV05} the analogous formula for $c_1,c_3,c_5$ in the case of a rectangle in place of the disk, and also for $c_7,c_9, \dots$ in the case of a sufficiently small rectangle
 were given, while \cite[Section 8]{SV05} ends by comparing with Siegel's mean value formula analogous to our Remark \ref{Rm2a}.
\end{remark}

\begin{figure}[t]
 \includegraphics[scale=0.8]{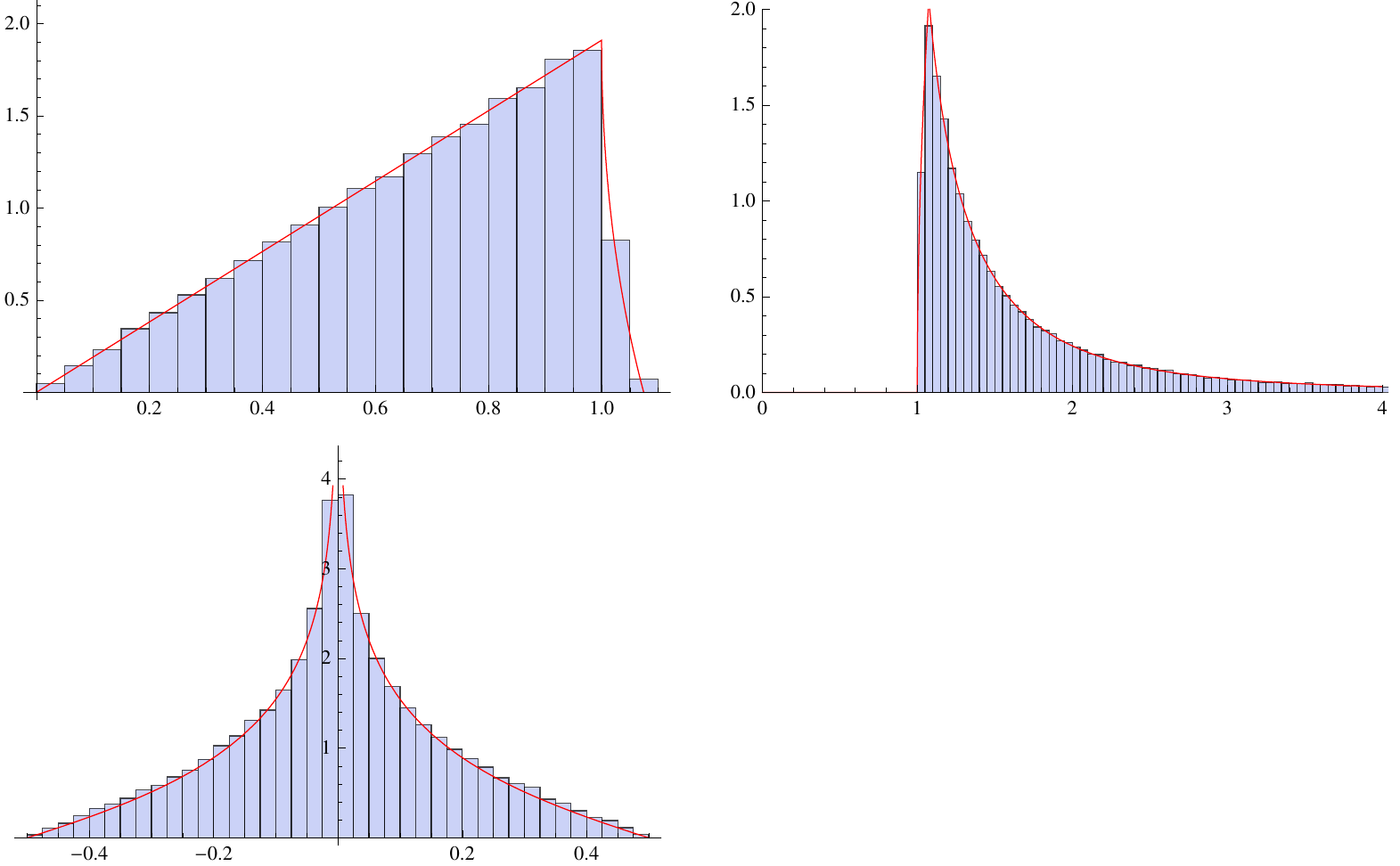} 
 \caption{\label{f2a} Numerically generated histograms for the distribution of the length of the shortest vector, the length of the second shortest linearly independent vector, and the cosine of the angle between these vectors for Haar distributed matrices in the fundamental domain, obtained by applying Lagrange--Gauss lattice reduction to $10^5$ Haar distributed elements from
 Sl$(2,\mathbb R)$ with larges singular value less than 100. The red curves are the theoretical predictions from Proposition
\ref{p4.2} }
\end{figure}

We would like to illustrate the results of Proposition \ref{p4.2} by first generating matrices
from SL${}_2(\mathbb R)$ with Haar measure, and then using Lagrange--Gauss
reduction of the corresponding lattice to the fundamental domain.  We generate
the matrices in the form of their singular value decomposition (\ref{MR}), with
$O_1$ and $O_2$ chosen with Haar measure from O$(N)$, and
the singular values generated according to the method of \S \ref{sS2}. The matrices
from  O$(N)$ can be generated by converting to Gram--Schmidt form the columns
of an $N \times N$ matrix of independent standard real Gaussians. In the case
$N=2$, the result of Proposition \ref{p4.1} tells us how to generate the singular values,
provided the largest singular value is no bigger than
$R$. For each matrix $M$ so generated, the Lagrange--Gauss algorithm (see 
e.g.~\cite{Br12}) is applied so as to reduce, using elements of
${\rm SL}_2^\pm(\mathbb Z)$, the column vectors of $M$ down to the fundamental domain. This is a simple and efficient task. Each $M$ can be viewed as consisting of two column vectors.
To initialise the algorithm, let $\mathbf u$ denote the shortest, and $\mathbf v$ the longest column vector. Step 1 is to calculate the scalar $\alpha = \lfloor (\mathbf u \cdot \mathbf v)/
||\mathbf u||^2 \rceil$, with $ \lfloor \cdot \rceil$ denoting the closest integer
function, and from this define the vector $\mathbf r = \mathbf v - \alpha \mathbf
u$. Step 2 is to update the shortest and longest vectors by defining
$\mathbf v := \mathbf u$, $\mathbf u := \mathbf r$. If indeed $||\mathbf u|| <
|| \mathbf v||$, steps 1 and 2 are repeated. If not, the process  ends and
returns the final updated values of $(\mathbf v, \mathbf u$) as the columns of $M$
reduced to the fundamental domain, with the first column corresponding to the lattice
vector with the shortest length. It is known (see e.g.~\cite{NS04}) that the total number of
steps required is bounded by a constant times the square of the logarithm of the 
longest length vector in $M$. Repeating this process many times allows us to form
histograms approximating the distribution of the shortest and longest basis vectors, and the
cosine of the angle between them. The results are displayed in Figure \ref{f2a}, showing excellent agreement between the theoretical and simulated distributions.

\subsection{The case $N=3$}
As written, the conditions (\ref{bb1}) for a Minkowski reduced basis in the case $N=3$
consist of an infinite number of inequalities. It was proved by Minkowski himself that
in fact a finite number of equalities suffice, the explicit form of which can be found
in \cite[\S 4.4.3]{Te88a} for example. On the other hand, it does not seem possible
to carry out the integrations needed to compute the exact form of the distributions
of the lengths and pairwise angles of the basis vectors. Nonetheless the numerical
approach used above for $N=2$ can be generalised.

The first step is to use the Metropolis Monte Carlo algorithm as detailed in the text
below Proposition \ref{p4.1} to generate the singular values of matrices from
 SL${}_3(\mathbb R)$ with Haar measure and bounded norm. Matrices $M$ from  SL${}_3(\mathbb R)$ with Haar measure can then be generated by using (\ref{MR}), as discussed in the second
 sentence of the paragraph below Remark  \ref{Rm1}. The task of transforming the columns of $M$  in the case $N=3$ to a Minkowski reduced basis  can be carried out using
 an algorithm due to Semaev \cite{Se01}. As input are three basis vectors 
 $\mathbf b_1$, $\mathbf b_2$, $\mathbf b_3$, ordered so that
 $|\mathbf b_1| \le |\mathbf b_2| \le |\mathbf b_3|$. Step 1 applies the 
 Lagrange--Gauss algorithm to $\mathbf b_1,\mathbf b_2$ and updates the vectors accordingly.  With $C = 1 - (\mathbf b_1 \cdot \mathbf b_2)^2/(||\mathbf
 b_1||^2 ||\mathbf
 b_2||^2)$ and
 $$
 x_2 := -  \Big \lfloor {1 \over C} \Big (
 {\mathbf b_2 \cdot \mathbf b_3 \over || \mathbf b_2||^2} -
  {\mathbf b_1 \cdot \mathbf b_2 \over || \mathbf b_2||^2} 
  {\mathbf b_1 \cdot \mathbf b_3 \over || \mathbf b_1||^2} \Big ) \Big \rceil, \quad
   x_1 := -  \Big \lfloor {1 \over C} \Big (
 {\mathbf b_1 \cdot \mathbf b_3 \over || \mathbf b_1||^2} -
  {\mathbf b_1 \cdot \mathbf b_2 \over || \mathbf b_1||^2} 
  {\mathbf b_2 \cdot \mathbf b_3 \over || \mathbf b_2||^2}  \Big ) \Big \rceil,
  $$  
  for step 2 set $\mathbf a = \mathbf b_3 + x_2 \mathbf b_2 + x_1 \mathbf b_1$.
Finally, in step 3, the process terminates if $||\mathbf a || \ge || \mathbf b_3||$.
Otherwise, $\mathbf b_3$ is replaced by $\mathbf a$, the updated
vectors $\mathbf b_1$, $\mathbf b_2$, $\mathbf b_3$ are ordered as in the input,
and the algorithm returns to step 1. It is proved in \cite{Se01} that the total number
of steps required is bounded by a constant times $\log (||\mathbf b_3||/||\mathbf v_1||) + 1)
\log || \mathbf b_3||$, where $\mathbf v_1$ denotes the shortest vector in the reduced
basis.

\begin{figure}[t]
 \includegraphics[scale=0.8]{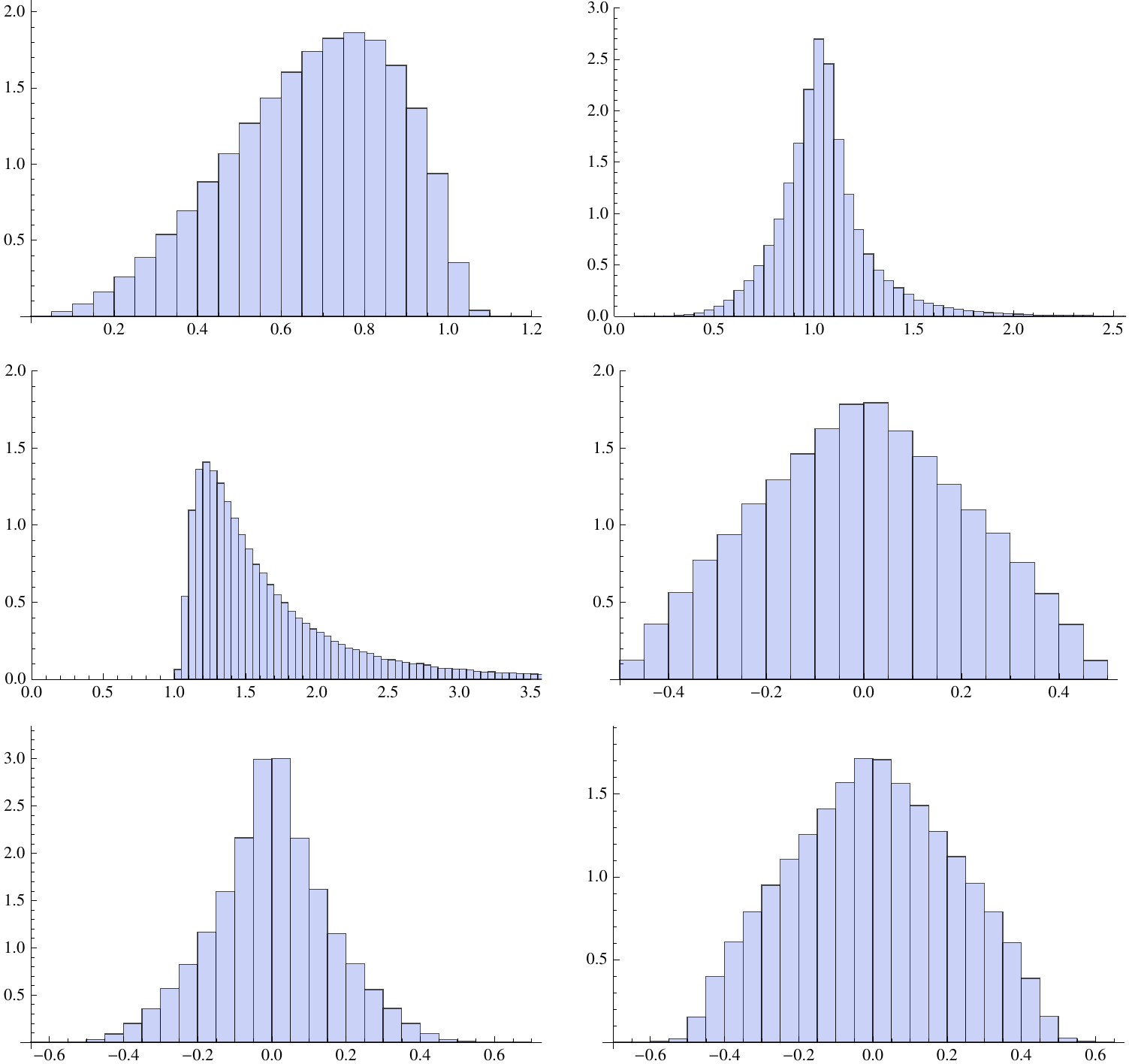} 
 \caption{\label{f2b} We denote by $\mathbf v_1$, $\mathbf v_2$, $\mathbf v_3$,
 the three Minkowski reduced basis vectors corresponding to a Haar distributed
 element of  SL${}_3(\mathbb R)$  and largest singular value
 bounded by $R=100$. The histrograms then correspond to 
  the distribution of the length of  $\mathbf v_1$, $\mathbf v_2$, $\mathbf v_3$,
  and the cosines of the angle between the pairs $(\mathbf v_1, \mathbf v_2)$,
  $(\mathbf v_1, \mathbf v_3)$ and $(\mathbf v_2, \mathbf v_3)$ respectively,
  with the vectors as generated by the procedure detailed in the text.
  }
\end{figure}

Implementing this procedure allows us to efficiently generate a large number
of Minkowski reduced basis vectors in $\mathbf R^3$ with Haar measure ---
which correspond to vectors with the lengths equal to the first three
successive minima --- and to
form histograms approximating the distribution of the lengths of these vectors,
and the cosines of their pairwise angles; see Figure \ref{f2b}. It appears in
the graphs that the largest permitted value of the shortest vector is, as for the $N=2$
case, $(4/3)^{1/4}$, which is in keeping with the face centred cubic lattice --- viewed
as alternate layers of hexagonal lattices ---
giving the most efficient packing of spheres.
 The smallest permitted value of the second shortest linearly independent vector lies
in the interval $(0,3,0,35)$, while again as for the $N=2$ case, the third shortest 
linearly independent vector
has shortest allowed length of 1 (corresponding to the simple cubic lattice). The cosine of the angle
between the shortest and second shortest basis vectors has magnitude less than or equal to
$1/2$, as for $N=2$, while this magnitude for the shortest and third shortest pair, and the
second and third shortest pair appears to be less than or equal to $1/\sqrt{3}$.
It remains as a challenge to further quantify these observation, and moreover to give an analytic description of the distributions.

One front on which such progress can be made is in relation to the small distance form of the
probability density function ,$p_1(s)$, say for the shortest basis vector. In the notation of 
Remark \ref{Rm2a}, for $N=3$ Siegel's mean value theorem tells us that $\Omega(R) = {4 \over 3}
\pi R^3$. On the other hand, trialling $p_1(s) = C s^2$ for $s$ smaller than the minimum allowed value
of the second smallest basis vector gives, according to reasoning of (\ref{w1})
$$
\Omega(R) = 2 C \int_0^R 
\Big \lfloor {R \over s} \Big \rfloor s^2 \, ds =
2 C R^3  \int_0^1 \Big \lfloor {1 \over s} \Big \rfloor s^2 \, ds.
$$
 Evaluating the integral according to the method of (\ref{w2}) shows 
 $\Omega(R) = 2 C \zeta(3) R^3/3$ and thus $C = 2 \pi/ \zeta(3)$. The functional form
 $p_1(s) = 2 \pi s^3/  \zeta(3)$ gives seemingly perfect agreement with the first histogram
 of Figure \ref{f2b} in the range $0\le s \le \mu$, for $\mu \approx 1/3$.
 
 We remark that the invariant measure on the space of unimodular lattices for $N=3$
 plays a fundamental role in the studies \cite{MS10,MS11} relating to the periodic
 Lorenz gas.

\subsection{The $N \to \infty$ limit}
The $N=3$ lattice reduction algorithm of Semaev \cite{Se01} has been described
in \cite{NS04} as a greedy version of two-dimensional Lagrange--Gauss lattice
reduction --- it used reduced vectors in dimension $N-1$ to obtain the
reduced basis in dimension $N$. However only for $N \le 4$ does the greedy
algorithm produce a Minkowski reduced basis \cite{NS04}. In higher dimensions this
latter task is both complicated and costly. Instead approximate lattice reduction is
used, with the best known method being the LLL algorithm, which guarantees
the shortest vector up to a factor bounded by $\beta^{(N-1)/2}$, $\beta \approx 4/3$.
Thus there
is a deterioration as $N$ gets large. On the other hand, it is in the limit $N \to \infty$
that an analytic description of the distribution of the shortest lattice vectors and
their pairwise angles again becomes possible for lattices corresponding to Haar
distributed SL${}_N(\mathbb R)$ matrices \cite{Ro55, So10, So11, Ki15, Ki16}.

Specifically, let $0 < \ell_1 \le \ell_2 \le \cdots$ denote the ordered sequence of the
lengths of the nonzero lattice vectors, with each pair $\pm \mathbf v$ counted as one.
Define $\nu_j := \pi^{N/2} l_j^N/\Gamma(N/2+1)$, which has the interpretation
as the volume of an $N$-dimensional ball of radius $\ell_j$. A result of
\cite{Ro55}, as generalised in \cite{So10,Ki16}, gives that with $k$ fixed and
$N \to \infty$, the sequence $\{\nu_l\}_{l=1}^k$ is distributed as a Poisson
process on $\mathbb R^+$ with intensity $1/2$. And with $\varphi_{jk}$,
$0 \le  \varphi_{jk} \le \pi/2$, denoting the
angle between the pairs of vectors with length $\ell_j$ and $\ell_k$, it is proved
in \cite{So11} that each $\tilde{\varphi}_{jk} := \sqrt{N}(\pi/2 - \varphi_{jk})$
has the distribution of the absolute value of a standard Gaussian random variable.

Generally the $N \to \infty$ limit of random lattices corresponding to
Haar distributed SL${}_N(\mathbb R)$ matrices is of interest from a number
of different perspective in mathematical physics; see e.g.~\cite{Ma00}.
The challenge suggested by the present work is to implement sufficiently accurate
lattice reduction in high enough dimension so that histograms analogous to those
of Figures \ref{f2a} and \ref{f2b} can be generated to illustrate the
results summarised in the previous paragraph.

\section*{Acknowledgements}
This research project is part of the program of study supported by the 
ARC Centre of Excellence for Mathematical \& Statistical Frontiers.
Additional partial support from the Australian Research Council through the grant DP140102613 is also acknowledged. Helpful and appreciated
remarks on an earlier draft of this
work have been made by J.~Marklof and A. Str\"ombergsson, with the latter
being responsible for the comments in Remark \ref{R4.6} relating to
\cite{SV05}.


\providecommand{\bysame}{\leavevmode\hbox to3em{\hrulefill}\thinspace}
\providecommand{\MR}{\relax\ifhmode\unskip\space\fi MR }
\providecommand{\MRhref}[2]{%
  \href{http://www.ams.org/mathscinet-getitem?mr=#1}{#2}
}
\providecommand{\href}[2]{#2}

\end{document}